\documentstyle[mnextra,astrobib,graphics]{mn-ab}
\seceqnum           
\def\ie{{\rm i.e. }}

\title[The Luminosity Function of Moderate
Redshift Clusters]
      {The Faint End of the Galaxy Luminosity Function in Moderate
Redshift Clusters}
\author[Wilson et al.]{Gillian Wilson,$^{\! 1,4}$ Ian Smail,$^{\! 1}$ 
Richard S.\ Ellis$^{2}$ \& Warrick J.\ Couch$^{3}$\\
\hbox{~}\\
$1)$ Department of Physics, University of Durham, South Rd., Durham DH1 3LE.\\
$2)$ Institute of Astronomy, Madingley Rd, Cambridge CB3 0HA.\\
$3)$ School of Physics, University of NSW, Sydney 2052, NSW Australia. \\
$4)$ Canadian Institute for Theoretical Astrophysics, University of
Toronto, 60 St. George St., Toronto, Canada M5S 1A7. \\}
\begin{document}

\date{{\sc Draft:} \today}

\label{firstpage}

\maketitle

\begin{abstract}
We present deep two-colour photometry of two rich clusters at $z=0.18$,
A665 and A1689.  We use these data to construct number counts as a
function of magnitude in the two fields.  By combining these counts
with similar observations from a large area field survey we
subtract the field contamination statistically to produce luminosity
functions for the two clusters.  Great care has been taken to achieve
agreement between the photometry of these two samples.  The cluster
data are complete to a limiting magnitude of $I=22.5$ or an absolute
magnitude in the cluster of $I= -18.0$ ($M^\star+5$).  The luminosity
functions of both clusters are well described by a Gaussian function
for the bright galaxies, combined with a Schechter function at the
faint end, similar to that required to fit the luminosity function in
local clusters.   The slope at the faint end of the Schechter function
in both clusters is extremely steep in $V$, $\alpha$~$\sim -2$.  A
shallower slope is seen to the limit of the $I$ data, indicating that
the cluster population is rapidly blueing as we reach fainter.  The
excellent agreement between the form of the luminosity function in our
two distant clusters, as well as agreement with the luminosity function given by
\citeN{driv-94} for a single $z=0.21$ cluster, indicates that this
faint blue population is a general constituent of distant clusters.  We
compare our results with those from studies of local clusters.  Depending
upon the degree of fading (or disruption) of these faint blue galaxies,
we tentatively identify their remnants with the low surface brightness dwarf
galaxies which are the dominant population in local clusters.  We
discuss the possible role of this population as the source of most
of the X-ray gas in rich clusters. 
\end{abstract}


\section{INTRODUCTION}

The form of the luminosity function (LF) of galaxies is a critical
constraint on models of galaxy formation \cite{cole-94}.  To measure
accurately the form of the galaxy luminosity function to faint
magnitudes requires the determination of distances to large numbers of
faint galaxies.  Observationally this is a very expensive procedure to
undertake for field galaxies, as the individual distances must be
measured spectroscopically.  However, by studying a rich cluster where
the bulk of the population is at a single distance, the practicalities
of LF estimation are optimised.  Unfortunately, this advantage comes at
a price; rich clusters are regions where we might expect environmental
effects on the form of the luminosity function to have their greatest
impact, evidence for which may already exist in the observed
morphology-density relationship \cite{dress-80,whit-93}.  Nevertheless,
by studying a number of clusters with a range of properties it may be
possible to gauge the effect of the environment on their galaxy
populations (itself an interesting issue) and hence determine the
likely similarity of field and cluster LFs.  It is for this reason that
much recent work has been concentrated on rich clusters.

A photographic study of 14 rich clusters by \citeN{col-87} and another 9
clusters by \citeN{lug-86} concluded that all clusters could be fit by
a universal LF with $\alpha\sim-1.25$ to $M_V\sim-19$\footnote{We adopt
$q_o=0.5$ and H$_o$ =50 km sec$^{-1}$ Mpc$^{-1}$.  With these
parameters 1 arcsec is equivalent to 3.91 kpc at $z=0.18$.}.  More
recent work has come to somewhat different conclusions.
\citeN{kash-95} carried out a photometric survey of both
morphologically-classified subsamples and the composite LF's of four
nearby clusters, including Coma. They found a faint end upturn which
led to a rejection of a single Schechter function with
$\alpha\sim-1.25$ as an adequate representation of the data; a
combination of a Gaussian at the bright end and a Schechter function at
the faint end being preferred. A faint end upturn in Coma was
originally seen by \citeN{abell-77} and later \citeN{met-83}  (See also
\citeN{thom-93}, \citeN{biv-95} and \citeN{seck-96}, all of whom also
prefer the combined Gaussian+Schechter function description of the LF).
However,  we note that \citeN{bern-95} found no upturn in their counts
in the {\it core} region of Coma.

In a seminal paper \citeN{bing-85} published the LF derived from their
extensive study of the galaxy population in the Virgo region.  The
advantage of working in the Virgo cluster is that its relative
proximity means that it is possible to reach far down the LF. Using
deep plate material they classified the galaxy populations
morphologically and then derived luminosity functions for various
types.  The brighter cluster ellipticals apparently follow a Gaussian
distribution, while the faint end of the LF is dominated by a
population of dwarf ellipticals which impart a steeply rising faint end
slope to the LF ($\alpha\sim-1.35$).  Later, \citeN{impey-88} (see also
\citeNP{both-91}) surveyed both Virgo and Fornax for low surface
brightness (LSB) objects and found many galaxies missed by
\shortciteN{bing-85}, steepening $\alpha$ yet further to $\sim-1.7$ in
Virgo (brighter than $M_V\sim -12$).  While the breakdown of the LF
into its separate morphological constituents was extremely
illuminating, this type of study can only be performed for nearby
clusters from the ground (although see \citeN{smail-96} for a study of
distant clusters using HST) and suffers from uncertain corrections for
field contamination.

A general picture is thus emerging of the form of the luminosity
function for local clusters.  The LF appears to be best described by
combining a Gaussian distribution for the bright (giant) galaxies with
a Schechter function for the fainter (dwarf) galaxies.  The very steep
slope of the faint component results in it being the dominant galactic
component in the clusters.  The evolution of this population is thus
critical for understanding, for example, how the blue star forming
galaxies seen in the cores of distant ($z\geq0.2$) Butcher-Oemler
clusters \cite{butch-84,oemler-92} have evolved by the present-day.
These galaxies should still populate such regions, although their
appearance will depend sensitively upon their subsequent star formation
histories.  One recently proposed mechanism for transforming them is
``galaxy harassment'' \cite{moore-96b}.  In this model star forming
field galaxies falling into the cluster potential have their dark halos
stripped, as well as undergoing tidal encounters with cluster
galaxies.  These processes remove the outer baryonic material from the
galaxy disk and concentrate its remaining gas in a nuclear starburst.
The end product of the process is suggested to be a dwarf spheroidal
galaxy.  Within the framework of such a model we would expect the
dwarf/giant spheroid ratio of the cluster to increase rapidly with
cosmic epoch, as increasing numbers of infalling galaxies are
transformed into dwarfs.

Another area of research which is sensitive to the form of the faint
end of the LF in distant clusters is the source of
intracluster gas and its enrichment.  \citeN{trent-94} has suggested
that a large fraction of the hot intracluster gas seen in Coma may have
originated in dwarf galaxies which, after an initial burst of star
formation, expelled their gas into the cluster potential and
subsequently faded to become the LSB dwarf galaxy population observed
today (or disrupted completely). To explain all the gas observed in
Coma, \citeANP{trent-94} concluded that an $\alpha\sim-1.8$ faint end
slope to the LF is required in the distant clusters, moreover this
steep population should be predominantly star forming dwarf galaxies.
It is therefore of considerable interest to determine observationally 
the form of the faint end of the galaxy luminosity function in distant
clusters.
  
\citeN{driv-94}[DPDMD] made the first attempt to measure the faint end
of the LF in a distant cluster ($z> 0.1$).  Their study showed evidence
for a steep slope to the LF in the range $M^\star+2$ to $M^\star+6$.
\citeN{driv-94} obtained a total integration time of 2.4 ksec in $B$
and $R$ of a $6 \times 4$ arcmin region in the distant cluster A963
($z=0.206$) using  the Hitchhiker parallel camera on the 4.2-m WHT.
They used the data to investigate the number counts in this region and,
by correcting statistically for the field contamination from counts
obtained with the same instrument in a number of blank fields, they
attempted to measure the cluster galaxy LF down to $R\sim 24$
(equivalent to $M^\star + 6$).  They found a steeply rising faint end
slope, with $\alpha \sim -1.8$ in a double Schechter parametrisation.
The slope of the faint end of the LF, (d$\log N$/d$m$) $\sim 0.3$, was
similar to the observed slope of the field counts.  Thus their result
could be reproduced by a simple zero-point shift  between the magnitude
scales of their field and cluster images, albeit of rather large
amplitude, $\delta m \sim 0.2$.  Unfortunately, owing to its
operational mode the Hitchhiker system was only calibrated once a year,
instrumental consistency has to be assumed, and thus it is difficult to
gauge the likelihood of such an error (They quote an rms scatter on the
zero point of $\pm 0.1$ mag).  To address this issue DPDMD compared
their photometry with data from a shallow photographic survey which
appeared to rule out any large offset.

To provide an independent check of the result reported by DPDMD we have
analysed deep two colour photometry of two distant $z=0.18$ clusters,
using a similar technique to measure the cluster galaxy luminosity
function.  Our cluster dataset is similar in quality and depth to
DPDMD's, although roughly a factor of six larger in area for each
cluster than that of DPDMD. To provide field counts with the same
selection criteria  and conditions as our cluster images, rather than
relying on published counts from the literature, we have analysed a
wide-field $V$ and $I$ imaging survey of blank fields.  In addition we
have taken great care to ensure the homogeneity of the photometric
systems used for both the cluster and comparison blank fields. 

We discuss our observational dataset in Section~\ref{sec:obs}, present
our analysis  in Section~\ref{sec:analysis}, discuss our
results in Section~\ref{sec:discussion} and give our main conclusions
in Section~\ref{sec:conclude}.  
 
\section{OBSERVATIONS AND REDUCTIONS}

\label{sec:obs}

\begin{figure}
\begin{center}
\scalebox{0.5}{
  \includegraphics[0,0][480,480]{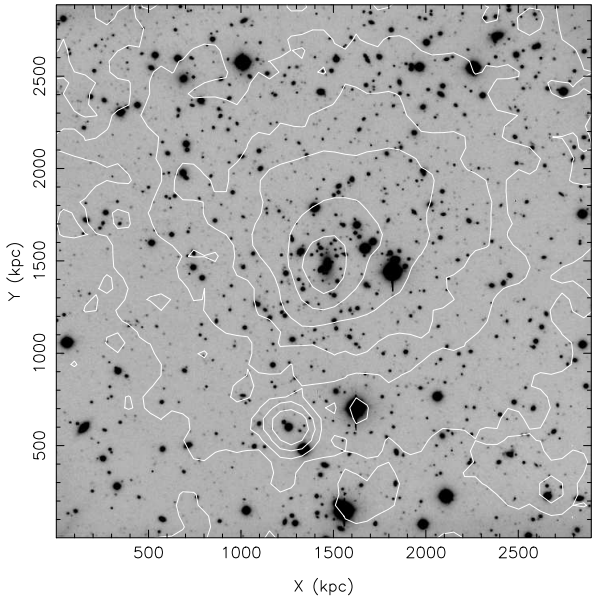}
}
\caption{An $I$ image of the rich cluster A665. The field of view is
$12.3 \times 12.3$ arcmin (2.9 $\times$ 2.9 Mpc). North is top and East
is left. Overlaid is a contour map of the X-ray emission from the
cluster.  The X-ray emission peaks on the central cluster galaxy and
shows an asymmetric distribution with a plume off to the north-west.
This structure is also visible in the distribution of red cluster
galaxies.
}
\label{fig:a665imap}
\end{center}
\end{figure}

\begin{figure}
\begin{center}
\scalebox{0.5}{
  \includegraphics[0,0][480,480]{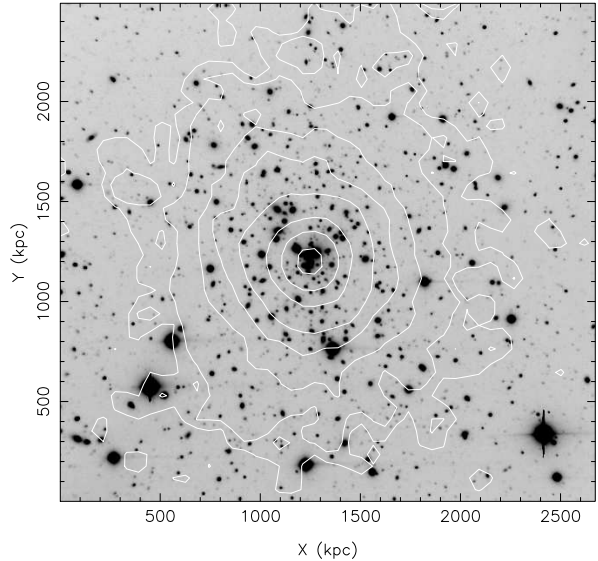}
}
\caption{The $I$ image of the cluster A1689. The field of view is $11.3
\times 10.6$ arcmin (2.6 $\times$ 2.5 Mpc) and North is top with East
to the left. Again we have overlaid the cluster X-ray distribution as a
contour map.  The highly symmetrical X-ray peak coincides with the
dense clump of bright galaxies which defines the cluster centre. 
}
\label{fig:a1689Imap}
\end{center}
\end{figure}

Before discussing our observations of the clusters and
the field we first describe the general characteristics of  
the two clusters observed.

\subsection {Cluster Properties}

The two clusters analysed in this study were A665 and A1689.  These
clusters both lie at $z \sim 0.18$ (A665 at $z= 0.182$ and A1689 at
$z=0.181$).  They are both optically rich with Abell richnesses of 5
and 4 respectively.   Using our imaging data we estimate the blue
fractions ($f_B$ from Butcher \& Oemler 1984) of both the clusters to
be low,  $f_B = -0.11\pm0.19$ for A665 and $f_B = 0.12\pm0.05$ for
A1689 (c.f.\ $f_B = 0.09\pm0.03$ for A1689 from Butcher \& Oemler
1984).  They have somewhat different morphologies, with A665 having a
single cD galaxy and A1689 containing a very dense central group of
galaxies (See Figs~\ref{fig:a665imap} and~\ref{fig:a1689Imap}).  In
the figures we have overlaid the ROSAT PSPC X-ray images taken from
the ROSAT archive to illustrate the general morphologies of the
clusters.  The two clusters are both strong X-ray sources with
luminosities in the 2--10~keV band of $L_X = 1.2 \times 10^{45}$ ergs
sec$^{-1}$ for A665 and $L_X = 2.0 \times 10^{45}$ ergs sec$^{-1}$ for
A1689 \cite{sh-83}.  In addition, both clusters have high velocity
dispersions: $\sigma_{cl} = 1201^{+183}_{-126}$ km sec$^{-1}$ (33
members) for A665 \cite{oeg-91} and $\sigma_{cl} = 1848\pm166$ km
sec$^{-1}$ (68 members) for A1689 \cite{gud-89}.   Finally, both
clusters show evidence of strongly lensed features in their core
regions \cite{tyson-90,kais-94b,tyson-95}. 

The characteristics of these clusters are very similar to the
cD-dominated cluster A963 studied by \citeN{driv-94}.  A963 is a
richness class 3 cluster at $z=0.206$ which is a strong X-ray source
($L_X = 0.95 \times 10^{45}$ ergs sec$^{-1}$, \citeNP{sh-83}) and also
contains two giant arcs \shortcite{lh-88,ellis-91}.   The cluster's
blue fraction has been estimated as $f_B = 0.19\pm0.05$ by Butcher \&
Oemler (1984).

\subsection{Data Acquisition}

The $V$ and $I$ imaging of the two clusters used in our analysis was
collected with the Prime Focus imager on the 2.5-m Isaac Newton
Telescope (INT), La Palma.  The comparison $V$ and $I$ observations of
blank fields needed to correct for field contamination come from deep
imaging with the f/1 camera on the 3.9-m Anglo-Australian Telescope
(AAT), Siding Springs (See \shortciteNP{lid-96} for more details).  In
addition we obtained additional photometric imaging to tie the
photometric systems of these two data sets together securely.  These
data were collected with the Prime Focus imager on the 4.2-m William
Herschel Telsescope (WHT), La Palma.

The imaging data on the two clusters, A665 and A1689, was collected
using a 2k$^2$ thick Ford CCD on the nights of 22--28
February 1993. Due to a technical problem with the data acquisition
system running the Ford chip a thick 1k$^2$ EEV CCD was used on the
night of the 23--24 February 1993.  The seeing during this run varied
between 1.4--2.3 arcsec, far in excess of the limit required for the
intended programme of lensing observations.  Accordingly, the 
current project was executed as a backup programme.  This
involved imaging in $V$ and $I$ of two clusters at $z\sim 0.18$
to study their LFs over a wide-field.

Our observing technique was to take multiple exposures (each of
$\sim1000$s) of the clusters, dithering the telescope by $\sim$ 15
arcseconds between exposures. This reduced flatfielding errors and
allowed us to create a master flatfield as explained in the next
section.  Observations of standard stars from \citeN{land-92} were
interspersed with the science observations throughout each night.  The
standards were observed across a large range in airmass to provide
atmospheric extinctions and zero points for the various clusters and
passbands.  Furthermore, the colour range chosen for the standard stars
spans the same range as expected for the cluster galaxies allowing us
to determine colour-terms for the various detector and filter
combinations.   Two nights during the run  were photometric and
calibration of all passbands was made using these nights.   We also
corrected our magnitudes for galactic absorption (reddening) using
$A_{I}=0.50A_{B}$ and $A_{V}=0.75A_{B}$ with $A_{B}$ given by
\citeN{burs-84} (see Table~\ref{tab:INT}).

\subsection{Data Reduction}

The reduction of our data followed standard procedures for the analysis
of deep imaging.  In particular, we chose to create super-flats using
the data frames themselves to allow us to best match the
characteristics of the flatfields to the science images.  The reduction
was all performed with standard IRAF routines and the procedure in
detail was:

\begin{enumerate} 
\item The frames were bias subtracted and trimmed.
The median bias level was obtained from the overscan region
of the chip and subtracted off.  The images were then trimmed to
remove the overscan strips.
\item Initial flatfielding was performed using twilight flats.
\item FOCAS (\shortciteNP{valdes-83}) was used to detect bright objects in the frames.
These were then removed and replaced by sky values drawn from regions
around the objects.
\item For a given frame, all the other cleaned
frames in that passband taken on the same night were median combined
to create a super-flat.
\item The initially flatfielded frames were then flatfielded using the
super-flat.
\item The images were aligned and geometrically re-mapped using
sub-pixel sampling to a single
basis frame (centred on the cluster), and then combined
using a clipped average algorithm in the IRAF task IMCOMBINE
to give final images for each passband/night. 
\item All final exposures from the different nights,
in a given passband, were co-added to 
provide a single frame of each cluster in each passband.
\end{enumerate}

The characteristics of the final datasets are summarised in
Table~\ref{tab:INT}.  Our cluster data covers a region of approximately
$12 \times 12$ arcmin or $2.8 \times 2.8$ Mpc in each cluster to a
depth of $I \sim 22.5$ and $V \sim 24.0$.  These are both equivalent to
$M^\star + 5$ at the cluster redshift. The average seeing ranged from
1.7  to 2.1 arcsec FWHM (Table~\ref{tab:INT}).   We use the mean colour
of the elliptical galaxy population ($(V-I)\sim 1.5$) to determine the
colour correction to transform our magnitudes onto the Landolt system.  We
include a contribution to the final magnitude errors from differences
between the adopted colour correction and the true value arising
from the observed range of galaxy colours in our fields.  The final
zero point errors were $\Delta I=0.05$ and $\Delta V=0.05$, obtained
from a combination of extinction, colour and frame-to-frame errors.

\subsection{Cluster Galaxy Catalogues} 

\label{ssec:galcats}
Having acquired and reduced our images we next needed to analyse them
to provide catalogues of object positions, magnitudes and colours.  To
achieve this we used the SExtractor analysis package \cite{bert-96}.
This package detects objects using an isophotal threshold above the
local sky (after convolution with a filter function), cataloguing those
with areas exceeding a minimum value.  SExtractor then  deblends the
objects and produces a catalogue of their properties.  The object
catalogue created includes information on the object positions, shapes,
profiles and magnitudes.  Tests to find the best detection parameters
for SExtractor were run on small sections of both the $V$ and $I$
images of the final cluster frames. The final values we adopted are
shown in Table~\ref{tab:SEX}.  While we used SExtractor to detect
objects and determine their centroids, to compare the number counts in
our various fields we have chosen to use large-diameter aperture
photometry (6 arcsec diameter or 23.5 kpc at the cluster distance).
This photometry was performed with  IRAF's PHOT package and we discuss
it in more detail later.

We corrected the catalogues for incompleteness at the faint end by
means of simulations (c.f.\ \citeNP{smail-95}).  We artificially
generated a ``typical'' galaxy for each of our four cluster fields by
co-adding a large number of bright galaxies ($I\sim 20$). We then
scaled the flux in this ``typical'' object to obtain simulated galaxies
of different magnitudes.  We added 200 such galaxies of a fixed
magnitude at a time to each of our science frames. We then re-ran
SExtractor on the resulting frames and noted what percentage of the
additional simulated galaxies were being missed.  The simulations thus
include the effects of incompleteness due to both the faintness of the
galaxies and also merging with brighter objects.  Note that implicit in
these simulations is the assumption that, over our range of interest of
$1-3$ magnitudes, faint galaxies have similar scalelengths to their
brighter counterparts. Even if this were not the case, and fainter
galaxies were intrinsically smaller, the high values of seeing ($\sim2$
arcsec) encountered during these observations make it likely that all
objects of differing magnitudes would be smoothed to a similar size.
Table~\ref{tab:INT} gives the 50 per cent completeness limits of the
catalogues obtained from these simulations.

To correct the catalogues for stellar contamination we have used the
relation between image size and brightness to separate the stars and
galaxies (c.f. \citeNP{lid-96}).  In a plot of isophotal radius versus
magnitude two distinct populations are evident at bright magnitudes,
with the more compact objects at any magnitude being the stars.  For
each frame a line separating the populations was determined visually
and the objects more compact than this limit were classified as stars
and removed from the counts.  At faint magnitudes ($I\geq18$) the two
populations begin to merge and it is more difficult to identify the
stars. Hence it seems likely that some contamination will remain,
although the fraction of stellar interlopers is small.  We are
primarily interested in the faint end of the LF and owing to the
relatively shallow star number counts compared to galaxies the star
contamination becomes significantly reduced as the magnitude increases
(see fig.\ 4 of \citeNP{lid-96}). After star removal, the A665 and
A1689 $I$ catalogues and A665 and A1689 $V$ catalogues contained 1689,
2035, 1659 and 1615 objects above their respective 50 per cent
completeness limits.

Before discussing the field counts used to correct our
cluster observations we first discuss whether the SExtractor detection
parameters that we used might cause us to miss a population
of low surface brightness galaxies.

\subsection{Are Low Surface Brightness Galaxies Being Missed?}

CCD frames are less susceptible to missing low surface brightness
objects than photographic plates (see e.g. \citeNP{bern-95} and 
\citeNP{turn-93} for more discussion). Despite this, and
to test the sensitivity of our object detection algorithm to the
adopted surface brightness limit, we also catalogued our cluster frames
using a procedure similar to that used by DPDMD in their
analysis of A963.  We selected a very low surface brightness limit
for our object detection with the understanding that the
ensuing sample would be strongly contaminated by noise objects.
Rather than making an arbitrary choice for the detection threshold
we chose to vary its value in the range 1.0--2.0 $\sigma$ and
selected the level which produced the highest number of new detections.
This is not simply the lowest threshold, owing to the effects of
the deblending algorithm on merged galaxy images.

To determine the excess number of real objects not detected by our
standard procedure we first removed all those objects from the
low-threshold (LSB) $I$ catalogues which occurred in the standard
versions of the catalogues.  Next we used the fact that the reddest
galaxies expected at the cluster redshifts ($z=0.18$) will correspond
to the colours of the spheroidal populations (E/S0) of the cluster ($(V-I) \sim1.6$).  Thus any real cluster objects will have colours bluer
than $(V-I) \leq 1.6$ measured from the seeing-matched $V$ and $I$
images, or for our limiting magnitudes of $I=22.0$ and $I=22.5$ they
would have $V$ $\leq$ 23.5 and 24.0
respectively. 

The number of objects from the LSB catalogues which are brighter than
$I=22.0$ (22.5), are undetected using the standard algorithm, and yet
have $V$ magnitudes brighter than $V=23.5$ (24.0) was 50 in A665 (2.5
per cent of the whole population) and 33 in A1689 (1.4 per cent)
respectively.  These proportions make a  negligible difference to the
cluster LF discussed later and so we conclude that our standard object
algorithm is not significantly biasing the object catalogue against low
surface brightness cluster members, in so far as they are detectable in
our data. 

\subsection{AAT f/1 Field Imaging}

The blank field observations necessary to field-correct our cluster
counts come from a wide-field $V$ and $I$ CCD survey of the equatorial LDSS
fields at 10$^{\rm hrs}$ and 13$^{\rm hrs}$.  The detector was a 1k$^2$
Thomson CCD with 0.98 arcsec/pixel sampling, providing a large field of
view ($\sim 17 \times 17$ arcmin).  The data were all taken in photometric
conditions and in seeing comparable to that encountered for our cluster
observations.  These data were kindly supplied and reduced by Dr.\ C.\
Lidman.  Dr.\ Lidman also provided internal calibrations between the
various fields with estimated internal zero point errors of $\Delta
\sim 0.03$ (\shortciteNP{lid-96}).  

\subsection{Calibration of the Comparison Field Photometry}

This final imaging dataset was acquired to tie the AAT and INT frames
rigorously to a single photometric system.   While both catalogs claim
to be on the Landolt (1992), it appeared prudent to us to actually test
this claim.  For this purpose the INT cluster fields and a selection of
the AAT blank fields were re-observed with a 1k$^2$ thinned Tek-2 chip
on the Prime Focus Imaging Facility of the WHT.  The conditions were
photometric but the seeing was very poor  ($\simeq4$ arcsec).
Observations of \citeN{land-92} standard stars across a wide range in
airmasses were also taken to provide robust photometric
transformations.  The science frames were reduced in a standard manner
using twilight flatfields before being rebinned to either the INT (for
cluster) or AAT (for field) pixel scales. The INT and AAT data were
degraded to the same seeing as the WHT and  10 arcsec diameter aperture
magnitudes then measured from galaxies in the fields.  We determine
magnitude offsets for the AAT data relative to the INT cluster fields.
The AAT field data was deemed to have magnitude offsets of $\Delta
(I_{AAT} - I) =0.07\pm0.05$ and $\Delta (V_{AAT} - V) =0.14\pm0.07$
relative to the INT system.   The relatively large offset uncovered in
the $V$ band is disappointing given the claimed accuracy of the two
catalogs, but we reiterate that it was to check for exactly this sort
of problem that we undertook these observations.  The source of this
offset has proved difficult to track down.  One check on the relative
magnitude scales in the INT cluster data is available from the colours
of the E/S0 sequence, this has a mean colour of $(V-I)=1.51$ at
$V=17.2$ with a scatter of $\delta(V-I) = 0.05$ in both
clusters.  This is in reasonable agreement with the no evolution
predicted colour of $(V-I)=1.57$, indicating that the INT photometric
is likely to have offsets of $\leq 0.06$.

We applied corrections to the field catalogues to put the field and
cluster photometry all onto the INT system.  By performing this
calibration of our control fields we were able to ensure that the
relative zero point errors between our cluster and field photometry
were reduced to a minimum ($\sim 5$ per cent random errors).

\subsection{Field Catalogues} 

\begin{figure*}
\begin{center}
\scalebox{0.9}{
  \includegraphics[15,60][560,660]{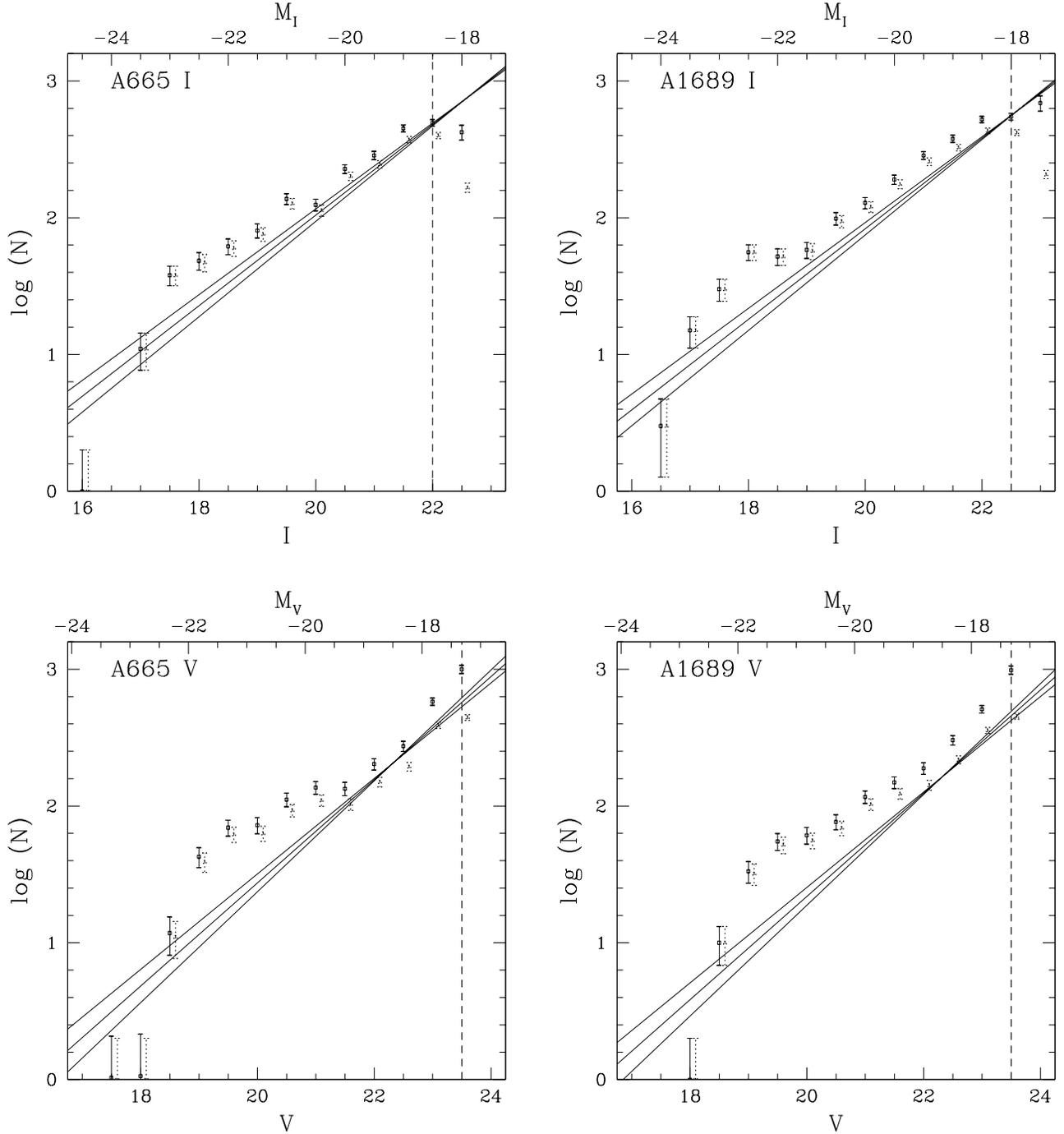}
}
\caption{Number counts for the $I$ (upper) and $V$ (lower) images of the two
clusters.  The counts are per field of view area for each cluster
(Table~\ref{tab:INT}) and per 0.5 magnitude bin. The solid points have
been corrected for incompleteness, the dotted points are the
uncorrected counts offset by 0.1 mag for clarity.  The equivalent field
counts from the AAT f/1 data are overlaid.  The three solid lines show
the best fit to the field counts and the  $\pm1\sigma$ errors. The dashed
line shows the 50 per cent incompleteness limit. Note the excess in the
cluster at $I\sim17$ and $V\sim18.5$ associated with the onset of the
bright cluster population. Stars have been removed from all the
counts. 
}
\label{fig:raw}
\end{center}
\end{figure*}

We analysed the AAT imaging data as in Section~\ref{ssec:galcats}, with
equivalent selection criteria. In our analysis we chose to use 9 $I$
fields (with seeing less than 2.1 arcsec) and 4 $V$ fields (with
seeing less than 2.6 arcsec).  We employed similar simulations to those
described in Section~\ref{ssec:galcats} to correct for incompleteness,
this time generating a typical galaxy for each passband (the observing
conditions and, in particular, the seeing for all the data frames taken
with each filter was approximately similar so it was only deemed
necessary to perform incompleteness simulations using one
representative frame in each passband).  The AAT field survey provided
adequate statistics for number counts as faint as $I\sim 22$ and $V\sim
22$, close to the limiting magnitudes of our cluster datasets.  Again,
we isolated and then removed the stars by plotting isophotal radius
versus magnitude.  

To measure the magnitudes of the objects on the cluster and comparison
fields we chose to use simple aperture magnitudes.  To provide
reasonable estimates of ``total'' magnitudes these were measured
in 6 arcsec diameter apertures centred on the object positions provided
in the SExtractor catalogues. The background sky was measured in
a wide surrounding annulus.  To remove differential aperture corrections
between the datasets we convolved all the images in a particular
passband to the seeing of the worst, using a Gaussian filter, before
measuring aperture photometry with IRAF's PHOT package.

In our analysis we have chosen to fit a single power law to the
observed field counts.  We used a maximum-likelihood technique, based
on minimising $\chi^{2}$ to determine the slope ($\gamma$) and
intercept ($C_m$) of the power law fit. The differential counts per
square degree per 0.5 magnitude were found to be:

\begin{equation}
\label{eq:fieldi}
\log_{10} dN  =  (0.331\pm{0.018}) I    - (3.24\pm{0.40})  
\end{equation}
\begin{equation}
\label{eq:fieldv}
\log_{10} dN   =  (0.377\pm{0.028}) V   - (4.73\pm{0.63})
\end{equation}

\noindent over the range 19.0--22.0 for $I$ and 19.5--22.0 for $V$
(the mean, and upper and lower limits on the counts are
shown by the solid lines in Fig.~\ref{fig:raw}).  The
errors in the slope and normalisation are composed of two components,
errors in the fit and frame-to-frame errors caused by Poissonian 
fluctuations between frames.  Our  slopes are
in reasonable agreement with those published previously by DPDMD
$\gamma_I = 0.34 \pm{0.03}$ ($I=19.0$--22.5) and 
$\gamma_V = 0.41 \pm{0.01}$ ($V=20.5$--23.0).
Furthermore, the number counts have been found to rise linearly in
these passbands to much fainter magnitudes (see e.g.\
\citeNP{smail-95}) and so we feel confident in marginally extrapolating
our field counts to obtain the same depth as used in the cluster
analysis. 

At this point we wish to quantify the effect gravitational lensing by
the clusters will have on the surface density of background galaxies
seen through the cluster.  Both clusters are known to be strong
gravitational lenses and they have the potential to increase or
decrease the observed surface densities of galaxies seen through them
\cite{broad-96}.  Two competing effect are present: the light
paths to distant galaxies are deflected by the cluster, resulting
in their images being displaced radially outward from the cluster
centre (effectively lowering the background surface density), meanwhile
faint galaxies are brightened by the lensing
amplification above our magnitude limit.
Using the methodology of \citeN{broad-96}, we expect 
the observed number density of galaxies,
N$^\prime$, to be related to the number density of galaxies in the absence
of the cluster, N$_{0}$, by:
\begin{equation}
\label{eq:lensing}
\frac{N^\prime}{N_{0}} =  A^{2.5s-1}  
\end{equation}

\noindent{where $A$ is the amplification and $s$ is the slope of the
field counts. Using a singular isothermal sphere approximation to the
cluster potential and our $V$ and $I$ band field count slopes of 0.377
and 0.331, we would expect to find a deficit of 2.5 per cent and 7 per
cent respectively in the background population in each passband for our
fieldsize.  These effects are negligible compared with the field-to-field
scatter and so we will ignore lensing in the ensuing discussion.}

\section{ANALYSIS}

\label{sec:analysis}

In summary, we have measured the differential galaxy counts in our
cluster fields and we have discussed the photometric zero point offsets
applied to the AAT catalogues to ensure consistency of photometry between
the AAT and INT data sets.  We have determined power law fits to the
background field counts and hence we are now in a position to
subtract the number of background galaxies statistically as a function
of magnitude from the total number of galaxies in our cluster fields,
to leave only the cluster members.  In the following sections we will
firstly fit single Schechter functions to our cluster luminosity
functions as is customary, before going on to show that improved
fits may be obtained using combined Gaussian+Schechter functions.

\begin{figure*}
\begin{center}
\scalebox{0.9}{
  \includegraphics[15,60][560,660]{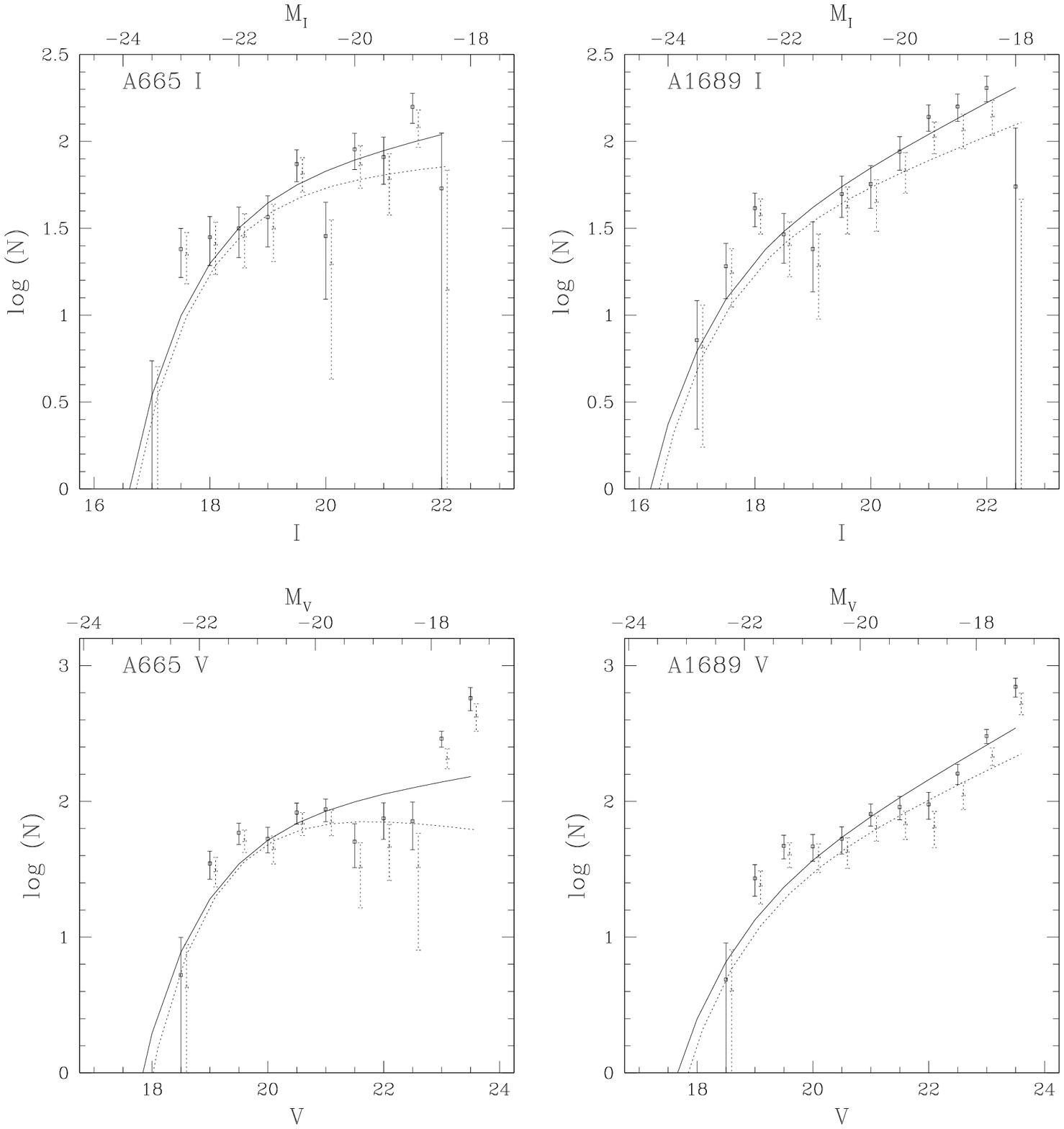}
}
\caption{Field corrected cluster LFs, defined as number of galaxies
per field per 0.5 mag.  The solid points and line have been corrected
for both incompleteness and obscuration by brighter objects (see text),
while the dotted points and line are only corrected for incompleteness
and have been offset by 0.1 mag for clarity.  Overlaid are the best fit
Schechter functions for each cluster.  The parameters are given in
Tables~\ref{tab:fieldins} and~\ref{tab:fieldinars}.
}
\label{fig:field}
\end{center}
\end{figure*}
\begin{figure*}
\begin{center}
\scalebox{0.9}{
  \includegraphics[15,60][560,660]{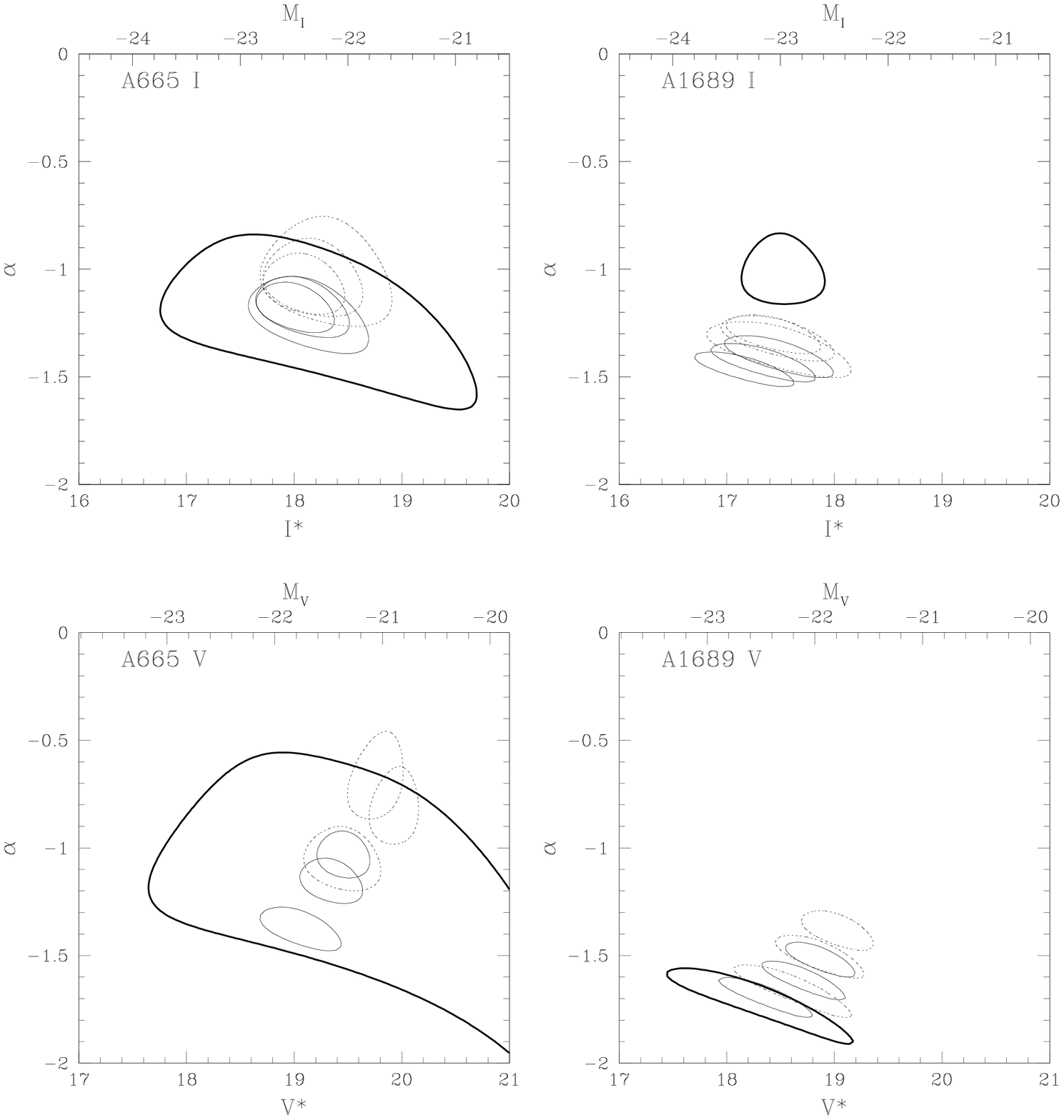}
}
\caption{$\alpha$ and $M^{\star}$ error contours. The contours are
$\chi_{\rm min}^{2}+6.3$ which corresponds to a $90$ per cent confidence
ellipse for normally-distributed errors and three free parameters.
The solid lines are from a Schechter fit to the incompleteness and
obscuration corrected points.  There are three lines corresponding to
the best and extreme ($\pm1\sigma$) values of the power law fits 
to the field counts. The dotted lines are from a Schechter
fit to the incompleteness only corrected points.   The 
bold line is from a Schechter fit to the
incompleteness and obscuration corrected inner/outer test points. 
}
\label{fig:cont}
\end{center}
\end{figure*}

\subsection{Measuring the Luminosity Functions}

We show the field-corrected cluster LFs for the clusters in both
passbands in Fig.~\ref{fig:field}. These include corrections for
incompleteness in the cluster frames.  One additional systematic effect
remained which had to be corrected for:  in a crowded cluster field there
is less open area available in which to detect the fainter galaxies
than in the corresponding field sample and so some faint background
objects will be obscured by brighter galaxies (as opposed to just
having merged isophotes).  This effect will also occur in the field but
will be less important.  We attempted to correct for this by computing
the available clear field of view for each magnitude interval by
subtracting from the total frame area the area covered by brighter
galaxies.  This technique will slightly overestimate the compensation
factor required, because it assumes that {\it all} faint objects which
happen to occupy the same region of sky as a brighter counterpart will
be obscured \ie that the bright galaxies are ``optically-thick'' at their
isophotal diameter.  The ``true'' obscuration correction will depend on
the surface brightness distributions in the bright and faint galaxies,
the internal extinction distribution in the bright galaxies and the
properties of the detection algorithm.  We claim, therefore,
that the two limits we use (no correction and the optically-thick case)
span the likely range.

As for the field counts, we again used a maximum-likelihood
method based on chi-squared to measure the $\alpha$, $M^{\star}$ and
$N^{\star}$ parameters in the Schechter function,
\begin{equation}
\label{eq:SchechtM2}
N(M)\,dM=kN^{\star}
e^{k(\alpha+1)(M^{\star}-M)}
e^{-e^{k(M^{\star}-M)}}\,dM
\end{equation}

\noindent where  $k=\frac{2}{5}\times(\ln10)$.

The generalised chi-squared statistic used was 
\begin{equation}
\label{eq:chi}
\chi^{2}= \sum_{{\rm i}} \left( \frac{\log_{10}N_{O}(\rm
i)-\log_{10}N_{E}(\rm i)}{\sigma(\rm i)} \right)^{2},
\end{equation}

\noindent where $N_{E}(\rm i)$ is the number of galaxies expected from
the Schechter function in the i'th bin, and $N_{O}(\rm i)$ is the
number observed. $\sigma(\rm i)$ is the  error for each bin, namely,
the Poisson error in the cluster-plus-field counts and the Poisson
error in the field counts, added in quadrature. In addition, we have
chosen to double the error for the final bin to reflect the greater
uncertainties associated with final point measurements and correction
values.

As a simple estimate of the scatter in the Schechter fit due to
uncertainties in the field counts, we fit for our Schechter parameters
three times, subtracting firstly the field counts derived from our best
fit field slope and normalisation parameters, and then the extreme
($\pm1\sigma$ values) (from equations~\ref{eq:fieldi}
and~\ref{eq:fieldv}).  This gives three best fit values for the
three cases.  As the Schechter function involves three parameters, the
confidence contours are three-dimensional shapes. The dotted lines in
Fig.~\ref{fig:cont} show one slice through each 3-D contour shape
($N^{\star}$ is held constant at its most likely value).  Ellipses
corresponding to $\chi^{2}_{\rm min}$+6.3 (the 90 per cent confidence
level for normally distributed errors with three degrees of freedom)
error contour are marked.   The most likely values of the Schechter
parameters are shown in Table~\ref{tab:fieldins}.  The function itself
is shown in Fig.~\ref{fig:field} (solid for both incompleteness and obscuration corrected
versions
and
dotted  for incompleteness
corrected only). To convert from apparent to absolute magnitudes we used
$M=m-5\log_{10}D-25-K$ where $D$ is the luminosity distance to the
cluster ($H_{0}=50$ kms$^{-1}$Mpc$^{-1}$) and K is the K-correction for
E/S0 galaxies (calculated by convolving our filter responses with the
spectral energy distribution of a present-day elliptical galaxy
redshifted to $z=0.18$).  The combined K-correction and distance
modulus, $m-M$, is 40.50 in $I$ and 40.82 in $V$.

\subsection{Independent Comparison Check}

We next determined a rough estimate of the cluster LF which was entirely
independent of the relative magnitude scales of the cluster and field
data.  Here we took advantage of the large field of view available in
our clusters and the strongly peaked distribution expected for galaxies
bound to the cluster potential.  We split our cluster frames into two
independent radial bins, each covering the same chip area.  For the
various images this translated into radii of $\sim 0$--0.9 Mpc and
$\sim 0.9$--1.3 Mpc.  We then simply subtracted the differential counts
in the outer region from those in the inner one.  This would completely
remove any galaxy population which has a flat distribution across the
frame, \ie the field population, and leave only the peaked cluster
galaxies.  In the absence of luminosity segregation within the clusters,
the resulting magnitude distribution of galaxies would represent the
global LF of the cluster.\footnote{We have tested for luminosity
segregation
\cite{lobo-96}
by combining the LFs in the inner and outer regions of the
two clusters and find only weak evidence for a steepening of the
faint end slope
$\alpha$ with radius in the clusters.}

\begin{figure*}
\begin{center}
\scalebox{0.9}{
  \includegraphics[15,60][560,660]{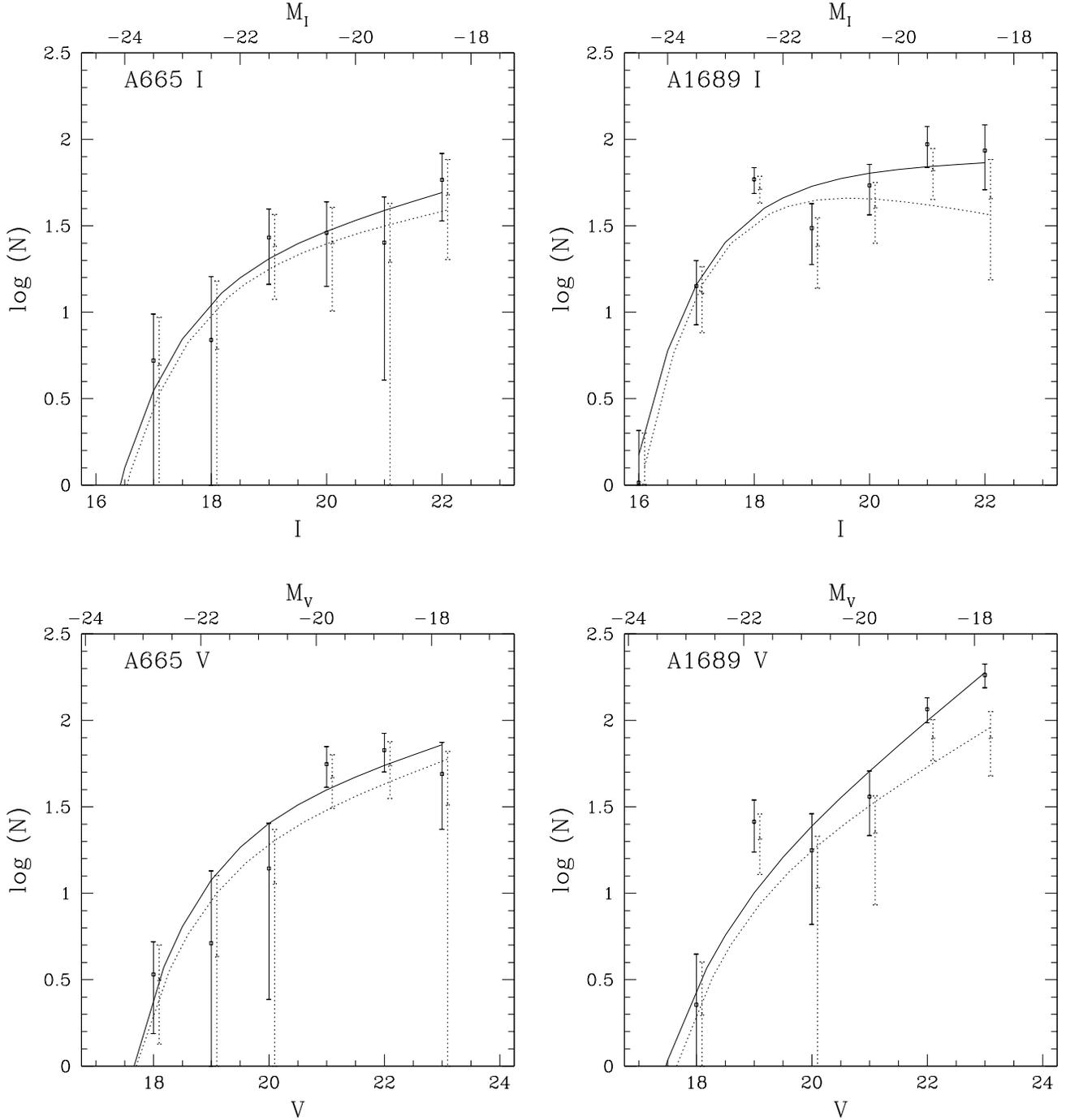}
}
\caption{Estimates of the cluster LFs obtained from
differential counts between the central and outer regions of each
cluster.  The solid points and line have been corrected for both
incompleteness and obscuration by brighter objects (see text). The
dotted points and line have been corrected for incompleteness
only 
and offset by 0.1 mag for clarity.  Overlaid are the best fit Schechter
functions for each case. The parameters for the incompleteness and
obscuration corrected fits are given in
Table~\ref{tab:ioinars}. 
}
\label{fig:io}
\end{center}
\end{figure*}

We show the results of this analysis in Fig.~\ref{fig:io}. The solid
points and line have been corrected for both incompleteness and
obscuration by brighter objects, while the dotted points and line have
only been corrected for incompleteness and are offset by 0.1 mag for
clarity.  Overlaid on Fig.~\ref{fig:io} is the best fit Schechter
function. The parameters are given in Table~\ref{tab:ioinars}. Note
that we might expect to obtain a smaller value for $N^{\star}$, the
normalisation, using this method, compared to the value obtained in the
field subtraction case. This is because some fraction of cluster
galaxies will be located in the outer annulus and will be subtracted
from the total along with the field galaxies. We see by comparing
values of $N^{\star}$ from Tables~\ref{tab:fieldinars} and
\ref{tab:ioinars} that this is indeed the case.

The bold lines in Fig.~\ref{fig:cont} show one slice through each
error fit to a Schechter function for the incompleteness and area
corrected points.  The errors are much larger for this test than the
field-subtraction one, but we reiterate that as a differential test
this is not affected by a zero-point magnitude error and is sensitive
to any population of clustered galaxies centred on the cluster centre.  

\section{RESULTS AND DISCUSSION}

\label{sec:discussion}

\begin{figure*}
\begin{center}
\scalebox{0.9}{
  \includegraphics[20,60][560,360]{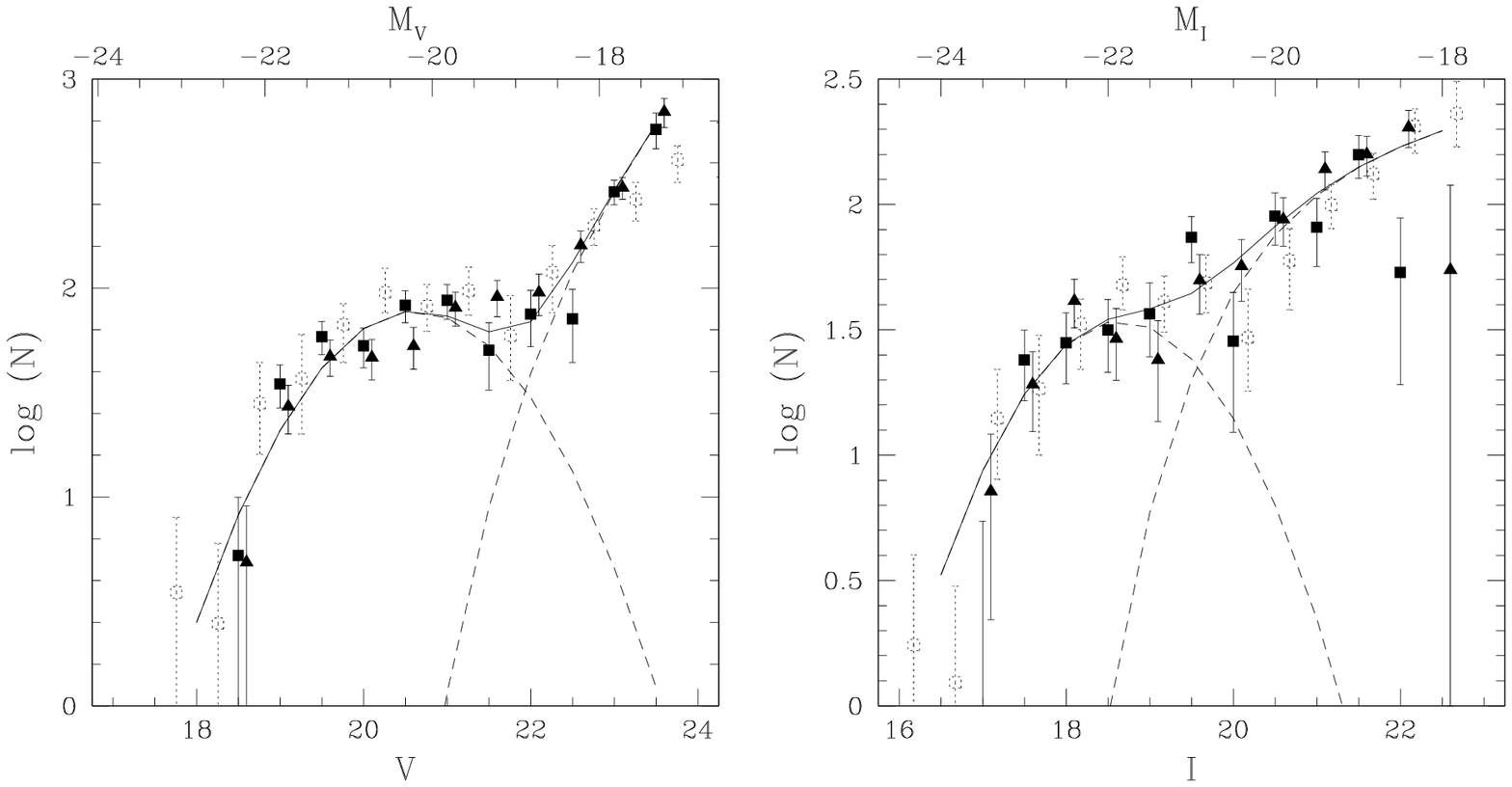}
}
\caption{Incompleteness and obscuration corrected points for A665
(solid squares) and A1689 (solid triangles). Also shown (open squares) 
is the data from Driver et al (1994)
which has been scaled as described in the text. Overlaid to our
data is the best fit Gaussian+Schechter function (See Table~\ref{tab:gs}
for the parameters). The A1689 data and line has been offset by 0.1 mag.
Note the good agreement between the form of the LF in
the three clusters.
}
\label{fig:driv2}
\end{center}
\end{figure*}

In Fig.~\ref{fig:field} we showed the best fit Schechter functions
including both incompleteness and area corrections.  In
Fig.~\ref{fig:io} we showed the results of an independent, if less
sensitive  test.  In Fig.~\ref{fig:cont} we made a quantitative
comparison of the various fits by displaying the error contours
associated with each of the distributions.  Comparing the A665 and
A1689 $V$ and $I$ error distributions in Fig.~\ref{fig:cont} with
each other it is clear that the best fit parameters for a single
Schechter function are only in very rough agreement, both between the
clusters and for the various fits within each cluster.  This is
surprising given the apparent similarity of the LFs in
Fig.~\ref{fig:field}.  The reason for this is apparent when we compare
the best fit Schechter functions to the data, especially the $V$ data.
Both clusters exhibit a faint component which rises very rapidly at
faint limits (The coarser binning and lower signal to noise in the
inner/outer comparison smears out this component and results in a
generally steeper faint end slope to the fits).  A single Schechter
function is incapable of simultaneously fitting both this feature and
the bright end of the LF, as is shown by the reduced $\chi^{2}/\nu$
values for the fits listed in Table~\ref{tab:fieldinars}.  These range
between 2.1 and 6.6 and indicate that a single Schechter function is
a poor analytic fit to our data points.  The disagreement between the
fits for the two clusters thus arises from the combination of the
incorrect functional form for the LF and the different contributions of
the faint component to the fit due to the slightly different magnitude
limits of the datasets. 
 
In
order
to decide upon a better functional form to describe the observed LFs we
again plot our $V$ and $I$ data in Fig.~\ref{fig:driv2}.  The A1689
data has been displaced (solid triangles) by 0.1 mag from the A665 data
(solid squares). However,  it was not necessary to scale the data
vertically (i.e.\ both clusters are of approximately equal richness).
DPDMD's A963 $R$ data is also shown displaced and overlaid (open
squares). We have applied a combined $(V-R)$ colour and distance modulus
displacement of +0.51 for E/S0 galaxies ($(R-I)$ of $1.08$), and also
an empirical normalisation constant.  The very close similarity between
the shapes of the three functions is immediately apparent, with all
three functions exhibiting a Schechter function form at the bright end
and then evidence for an upturn at magnitudes fainter than $V\sim 22$
($I\sim 20$).  In the following discussion we will focus on the $V$
data where this upturn is more prominent.

In the paper of DPDMD it was also noted that the upturn at the faint end
ensured that a single Schechter function was not a satisfactory fit to
the data. DPDMD decided to fit a double Schechter function
to their data, fixing the slope of the first function (with the brighter
knee) to be equal to $-1.0$.  Here, instead, we elected to fit a
combination of a  Gaussian function at the
bright end of the form:
\begin{equation}
\label{eq:gauss}
N(M)=Ke^{-(M-M^{\rm cent})^{2}/2\sigma^{2}}
\end{equation}

\noindent{and a Schechter function given by
equation~\ref{eq:SchechtM2}
at the faint end. We prefer this description over the
double Schechter function as it has been used extensively in
investigations of the LFs of local clusters
(e.g. \citeNP{met-83,seck-96,thom-93,biv-95}), including morphological studies
(e.g. \shortciteNP{bing-85}) which have shown that giant galaxies tend
to follow a Gaussian distribution and dwarfs a Schechter function.

As there is some degeneracy between the various parameters we have elected to
fix the the peak of the Gaussian, $K$, to be 10 per cent of the
normalisation, $N^{\star}$, of the Schechter function (reasonable
values lie in the range 0.08 -- 0.14).  We also chose
to fix the standard deviation of the Gaussian to be $\sigma=1.0$, in
keeping with values determined locally for Coma 
\shortcite{met-83,biv-95,seck-96}.  Table~\ref{tab:gs} gives the best fit
parameters for the Gaussian+Schechter function description of the
individual cluster LFs, as well as those for the samples without the
obscuration correction.  As is apparent, this function provides a much
better description of the data than a single Schechter function. This
is supported by the reduced $\chi^{2}$ values for the
Gaussian+Schechter fits  which are all
significantly better than the single Schecter fits 
(Table~\ref{tab:fieldinars}) .  The reduced
$\chi^{2}$ values are still  higher than might be expected, although we
have not allowed for the effects of clustering and have low numbers of
objects in our bright bins.  Nevertheless, the Gaussian+Schechter fits
allows us to make a robust quantitative comparison of the form of the
LFs in the two clusters.

We see from Table~\ref{tab:gs} that there is remarkable consistency
between the $M^{\rm cent}$ and $M^{\star}$ values for each bandpass.
In the $V$ band, $M^{\rm cent}$ is $-$20.27 for A665 and $-$20.19 for
A1689, and in $V$ it is $-$22.14 for A665 and $-$22.06 for A1689.
There is also close agreement between the values of $M^{\star}$, with
$-$18.13 for A665 and $-$18.50 for A1689 in $V$, and $-$20.49 for A665
and $-$20.47 for A1689 in $I$.  In view of this good agreement between
the two clusters we therefore combined their LFs and show the fit to
this combined dataset in Fig.~\ref{fig:driv2}.  The parameters used
may be found in Table~\ref{tab:gs}.

The main point of interest in comparing the cluster luminosity
functions in $V$ and $I$ is the much steeper faint end slope in $V$
compared to $I$. This difference implies that fainter cluster galaxies
must have colours which are significantly bluer than the bright
ellipticals. In order to quantify this effect we plotted the cumulative
number of galaxies (using our best Gaussian+Schechter fits) in each
passband as a function of magnitude.  If both passbands were detecting
the same objects then comparing the magnitude limits in $V$ and $I$ as
a function of cumulative density would give the typical colours as a
function of magnitude in the population.  We found that the mean $(V-I)$
shifts from $(V-I)\sim 1.6$ at $I\sim 16$ (the colour of the bright
cluster ellipticals) to $(V-I)\sim 0.6$ by $I=22$
($M=-18.5$).   The steady blueing of the cluster elliptical
population as a function of magnitude, presumably due to metallicity
effects, amounted to only $d(V-I)/dI = 0.07$ per mag, and so can only
account for half of the colour shift.  Thus the difference in
the faint end slopes in the $V$ and $I$ passbands apparently reflects a
rapid blueing trend in the faint galaxy population in the clusters, with
the faintest cluster members having the colours and luminosities of
typical dIrr galaxies.
  
As we discussed earlier, the Gaussian+Schechter parameterisation has
been used to describe the LF of the Coma cluster (an Abell richness
class 2 cluster).  Here, we compare the properties of the bright galaxy
populations in our distant clusters, as described by the Gaussian
component, with those seen in Coma.  The  value of $M^{\rm cent}$ for
our combined cluster sample is $M^{\rm cent} = -20.20$ in $V$, very
close to that found for Coma: $-$20.03 \cite{thom-93}, $-$20.20
\cite{biv-95} and  $-$20.16 \cite{seck-96} (converting all measurements
to the $V$ band using $(B-V)=1.0$).  The scatter amongst the various analyses of Coma
indicates that we ought not to put too much weight on this comparison,
but it is interesting to note that the distant clusters appear to be
$\sim 0.1$ brighter in rest-frame $V$ compared to the local value. This
degree of brightening is not unreasonable out to $z=0.18$ in a
population of passively evolving elliptical galaxies formed at high
redshift \cite{barge-96}.

\begin{figure}
\begin{center}
\scalebox{0.55}{
  \includegraphics[35,30][500,500]{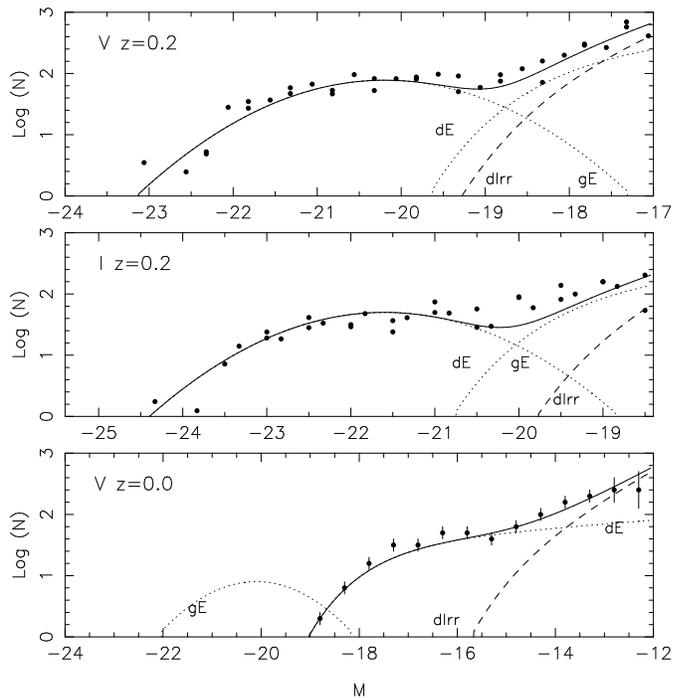}
}
\caption{Our simple model for LF evolution in clusters.  The composite LF
of the cluster is shown by the solid line (except for the $z=0$ case
where we show the composite dwarf LF), the various sub-components are
shown by dotted/dashed lines.  Overlaid on these are the observed data
points from Fig.~\ref{fig:driv2}.  We show in the bottom panel the
dwarf elliptical LF derived for Virgo by Impey et al.\ (1988) converted
into $V$ using $(B-V)=0.6$ (applicable to dEs), where we have 
increased the size of the
error bars at the faint end to take into account the uncertainties in
the corrections applied by Impey et al.
}
\label{fig:model}
\end{center}
\end{figure}

Turning to the faint cluster population, parameterised by the Schechter
function, we find that locally the values of the characteristic
luminosity $M^{\star}$ are more poorly defined and have larger errors;
$-$18.1 \citeA{biv-95},  $-$18.7 \citeA{seck-96} and $-$18.0 for
the dE population in the Virgo cluster \shortcite{bing-85}.  Again
these encompass the value observed in our distant clusters,
$M^{\star}=-18.5$, indicating a similar lack of strong evolution in
the characteristic luminosity of this population to that seen for the
brighter galaxies.  However, when we compare the faint end slope,
$\alpha$, for the faint cluster population we find a large difference
between the distant and local clusters.  The faint end slope in Coma
lies in the range $\sim-1.1\:$--$\:-1.4$ (\citeANP{biv-95} and
\citeANP{seck-96}) and possibly even steeper ($\alpha \sim -1.7$ for
the dE population of Virgo \shortcite{impey-88} ), compared to 
$\alpha \sim -2$ for our distant clusters
\footnote{The divergent slope found in our fits can easily be reduced by 
selecting a different characteristic luminosity in the
Schechter part of the LF.}.

The consistency of the upturn at the faint end in the three clusters
available to us, particularly in the $V$ band, suggests that this
feature is common to many distant rich clusters. The fact that the
upturn appears much less dramatic in the $I$ band indicates that faint
end population is significantly bluer than the typical bright cluster
galaxies.  To obtain an overview of the implications of these observations
we have constructed a simple model which takes the broad features of
the observed $z\sim0.2$ $V$ band LF discussed above and predicts both
the $z\sim0.2$ $I$ band LF and the evolved $V$ band LF which we would
expect to observe locally.  The key feature of this model is the
division of the faint component of the cluster LF into two equal
populations, dwarf ellipticals (dE) and dwarf irregulars (dIrr).  The
LFs for
both of these groups follow a Schechter form, but the different 
star formation histories expected for the two populations result
in significantly different evolution to the present day.

In the top panel of Fig.~\ref{fig:model} we replot our $V$ band data
from Fig.~\ref{fig:driv2}. We also show the constituent LFs of the
three populations assumed in our simple model, and the resultant LF.
The giant elliptical galaxies (gE) have been assigned the same $M^{\rm
cent}$ and $N^{\star}$ values as discussed in Table~\ref{tab:gs}. The
dwarf elliptical and dwarf irregular populations have equal
normalisations and characteristic luminosities ($M_V\sim -17.8$) but
have radically different faint end slopes. The dwarf ellipticals have
been assigned a value of $\alpha \sim -1.2$ (chosen because this was the
value observed for our $I$ band LF). The dwarf irregular population has
a much more steeply rising slope (with $\alpha \sim -2$). 

We next used the colours expected for these three populations to
predict the expected form of the $z\sim0.2$ $I$ band LF.
We used the observed colours of the giant and dwarf ellipticals
taken from fits to the E/S0 sequence in the clusters
and estimated at their characteristic magnitudes ($(V-I)\sim1.5$ for Es, 
and $(V-I)\sim1.2$ for dEs) to  transform their LFs from $V$ to $I$.
For the dIrr population we adopted the colour of a dIrr galaxy (close
to flat spectrum), $(V-I)\sim0.6$.  The three transformed LFs are shown in
Fig.~\ref{fig:model}, as is the expected composite $I$ band LF.
As can be seen this is in reasonable agreement with the observations
which we have overplotted.

Finally we used our model to predict what present day $V$ band LF we
might expect to observe in a rich cluster.  The two faint end
populations will evolve very differently with time.  The dwarf
ellipticals will evolve passively in a similar manner to the bright
cluster ellipticals (with fading by about 0.1 mags in rest-frame $V$
between $z=0.2$ and today). The very blue (star forming) dwarf
irregulars, however,  are assumed to have their star formation stopped
soon after we observe them (around $\sim 3$ Gyrs ago), either due to a
violent truncation mechanism (such as ram-pressure stripping of their
disks or a galactic wind) or by the slower removal of the
gas-reservoirs in their halos.  This population will then fade rapidly (by
upwards of 2--3 mags in rest-frame $V$ \cite{barge-96}, and possibly up
to $\sim5$mag \cite{babul-96}).  An even more drastic possibility would
be that the galaxies are partially or totally disrupted due to tidal
interactions in the cluster and either become low surface brightness
dwarfs (the blue and hence relatively young objects discovered in Virgo
by \shortciteN{impey-88} or be removed from the scene altogether
(c.f.\ \citeNP{babul-92}).  Whatever the mechanism, the dwarf
irregulars are assumed to have faded dramatically (3 mags) by the
present day.  We show the individual LFs in the bottom panel of
Fig.~\ref{fig:model} and also the composite dwarf LF (the faded dIrr+dE
population). We have also plotted in Fig.~\ref{fig:model} the
\shortciteANP{impey-88} LF of dwarf ellipticals in the Virgo cluster.
Clearly, this can be adequately described using our combined dwarf
population.  It is interesting to note also that the region around
$M_V\sim -16$ where the dIrr begin to appear in reasonable numbers is
also the point at which \shortciteANP{impey-88} report the breakdown in
the surface-brightness--luminosity correlation in their data,
indicating that these fainter objects probably arise from a very mixed
group of progenitors.

In summary, the significant difference in the colours of the dwarf
ellipticals and dwarf irregulars implies that the former dominate in the
$I$ band LF of the distant clusters to the limit of our data (although
not much fainter) and the latter determine the faint end in the $V$
band.  The differential evolution in the dwarf irregulars results in 
them fading far down the LF by the present day -- leaving the
intermediate regime $M_V \sim -18$ dominated by the dwarf ellipticals.
Below $M_V\sim -16$ remnants of the dwarf irregulars begin to be
seen in large numbers resulting in an upturn in the LF.  

Finally, we used our model to predict the early-type dwarf to giant
galaxy ratio (EDGR) in our distant clusters. Local studies \ie
\citeN{seck-96} find an EDGR of $1.8\pm0.6$ in the Coma cluster to a
limit of $M_B=-15.5$ ($M_V\sim-16.1$), similar to the value of 2.1
found for Virgo by \citeN{ferg-91}. Note that we compared our results
only with the brightest subsample from \citeANP{seck-96} and
\citeANP{ferg-91} because this required the least extrapolation beyond
the limit of our data.  If we calculate the EDGR from the dE and gE
populations in our model at $z\sim0.2$, we obtain an EDGR of 2.3 for
$(K/N^\star)=0.1$ (ranging from 3.3 for $(K/N^\star)=0.08$ to 1.9 for
$(K/N^\star)=0.14$).   Little evolution is expected in our model for
the EDGR brightward of $M_V\sim -16$  as the faded remnants of the
dwarf irregulars do not appear brighter than this at the present day
(It is not clear whether the proposed population of dwarf irregulars
could metamorphose into e.g.\ dwarf spheroidals  in the interim between
$z\sim0.2$ and the present day. The point we stress here is that with
our assumption of 3 mags fading the remnants cannot affect the EDGR
brightward of $M_V\sim -16$ whatever form they take).   Thus from our
model we expect an EDGR of 2.3$^{+1.0}_{-0.4}$ in local clusters,
consistent with the values observed.  We conclude that there is
currently no convincing evidence for strong evolution in EDGR out to
$z\sim 0.2$.  

Finally, turning to the ``galaxy harassment'' model of \citeN{moore-96b} we see
that the predicted evolution of the EDGR will depend critically upon
the relative effect of disk stripping and central starbursts on the
luminosity of the remnant dwarf spheroidal.  \citeANP{moore-96b} suggest
that the infalling field galaxies are somewhat brighter than the dIrr
population in our models ($M_V\sim -19$).  If the stellar luminosity
lost due to stripping is less than a few times that produced by the
starburst, the remnants would appear brighter than $M_V\sim -16$ in the
local clusters and hence the present day EDGR  would be higher
than that seen at intermediate redshift, in contradiction to the
observations.  Clearly there is considerable scope for obtaining
agreement between the harassment model and the observations, but we
are encouraged by the future possibilities of closely comparing theoretical
models with observations of galaxy evolution in rich clusters.

\section{CONCLUSIONS}

\label{sec:conclude}

\begin{enumerate}

\item We have presented deep $V$ and $I$ photometry of A665 and A1689,
two rich, X-ray luminous clusters at $z=0.18$.  This photometry reached
a limiting apparent magnitude of $I=22.5$, equivalent to an absolute
magnitude of $I=-18.0$, or $M^{\star}+5$, at the cluster redshifts.  We
analysed the data to provide differential number counts of
$\sim1500$--2000 galaxies in the fields of each of these two clusters.

\item We used an independent $V$ and $I$ field survey reaching
similar depths  to remove the field contamination from our cluster
fields.   We undertook extensive observations to calibrate the
photometry between the field and cluster datasets.  This gave us
confidence in our field subtraction,  and in the resulting cluster LF. 

\item The LFs we derived in our two clusters also showed remarkable 
consistency with each
other.  Both showed knees, a flat region and then an upturn at the
faint end (beginning around $M_V=-19$).  The upturn in the cluster LFs
was more marked in $V$ than $I$ ($\alpha \sim -2$ versus 
$\alpha \sim -1.2$) indicating that the faint population
rapidly becomes bluer with decreasing luminosity. By means of a simple
test using cumulative plots of the galaxies in each band we showed that
the mean colour was $(V-I)\sim 0.6$ by $I\sim 22$ in contrast with the
mean colour of the bright spheroid population, $(V-I)\sim 1.6$.
Moreover, the form of the LF in our two clusters was in very good
agreement with that found for a single distant cluster by
\citeN{driv-94}.

\item We constructed a simple model which described the gross
features of the LFs observed in our distant clusters.  The main feature
of this model was the partitioning of the faint cluster component into a
quiescent red dwarf elliptical population and a star forming blue dwarf
irregular population.  Evolving these populations forward to the
present day we could fit the form of the dwarf luminosity
function observed in Virgo if we assumed substantial, but not
unreasonable, fading in the dIrr population.

\item If the cessation of star formation and subsequent
fading/disruption of these blue dwarf galaxies is associated with
extensive gas loss from these systems, then the population seen in our
distant clusters has a sufficiently steep faint end slope ($\alpha \sim
-2$) to be the source of all the X-ray gas seen in local rich clusters \cite{trent-94}.

\item Extending observational studies such as the
one presented here to higher redshifts will provide further
constraints on the changing form of the giant and dwarf
galaxy populations in clusters, and hence the consequences for the
evolution of the X-ray emission of the clusters.
\end{enumerate}

\begin{table*}
\caption{\hfil INT cluster observations\hfil}
\vspace{0.5cm}
\begin{tabular}{lcccccccccc}
Target & Filter & Chip & Scale & Field & T$_{\rm exp}$
& FWHM & m$_{\rm lim}$ & $\mu (1\sigma)$ 
&  Reddening & N$_{\rm gal}$\\
& & & ($''/{\rm pix}$) & & (ksec) & ($''$) & (50$\%$)  & (mag$/\Box''$) & & $(<m_{50})$\\
A665  & $I$ & FORD & 0.37 & $12.34' \times 12.30'$  & 20.5 & 1.7 &
22.0& 25.3&0.007&1689\\
A1689 & $I$ & EEV  & 0.54 & $11.29' \times 10.61'$  &  9.8 & 1.8 & 22.5 & 26.1 & 0.063&2035 \\
A665  & $V$ & FORD & 0.37 & $12.34' \times 12.30'$  & 17.0 & 2.0 & 23.5 & 26.7 & 0.015&1659 \\
A1689 & $V$ & FORD & 0.37 & $11.29' \times 10.61'$  & 18.0 & 2.1 & 23.5 & 26.9 & 0.098&1615 \\
\end{tabular}
\label{tab:INT}
\end{table*}

\begin{table*}
\caption{SExtractor detection parameters}
\vspace{0.5cm}
\begin{tabular}{lcccc}
{Cluster} & {Filter} & Conv.\ & Min.\ Area & $\mu_{\rm thresh}$ \\
& & & (pixels) & (mag$/\Box''$) \\
A665 & $I$ &$3\times 3$ & 6 &24.5 \\
A1689 & $I$ &$3\times 3$ & 6 &25.1 \\
A665  & $V$ &$3\times 3$ & 6 &26.3 \\
A1689 & $V$ &$1\times 1$ & 6 &26.3 \\
\end{tabular}
\label{tab:SEX}
\end{table*}

\begin{table*}
\centering
\caption{Schechter function parameter fits for incompleteness
corrected field subtracted points. The columns show the cluster, filter,
faint end slope, apparent magnitude at the knee, absolute magnitude at
the knee, normalisation, $\chi^{2}$, and reduced $\chi^{2}$ values. The
best fit parameters found by subtracting the mean value (first row) and then
the $\pm1\sigma$ error values
of background counts (second and third rows) are also shown.}
\vspace{0.5cm}
\begin{tabular}{lccccrcc}
{Cluster} & {Filter} & $\alpha$ & $m^{\star}$ & $M^{\star}$ &
$N^{\star}$ & $\chi^{2}$ & $\chi^{2}$/$\nu$  \\
A665  & $I$ &   $-$1.065 &  18.11 & $-$22.39 & 127.15 & 24.67 & 2.74 \\ 
 &  & ${-1.083}$  & ${18.04}$ & ${-22.46}$ & 
${143.03}$ & ${24.54}$ \\
  &  &     ${-1.060}$  & ${18.18}$ & ${-22.32}$ & 
${103.57}$ & ${25.41}$ \\
A1689 & $I$ &   $-$1.321 &  17.42 & $-$23.08 & 62.63 & 21.34 & 2.13 \\
& & ${-1.303}$  & ${17.40}$ & ${-23.10}$ &  ${73.40}$ & ${22.40}$ \\
  &  &    ${-1.366}$  & ${17.40}$ & ${-23.10}$ & 
${46.76}$ & ${20.54}$ \\
A665  & $V$ &   $-$0.843 &  19.64 & $-$21.18 & 240.54 & 49.36 & 5.49 \\
& & ${-0.709}$  & ${19.73}$ & ${-21.09}$ & ${296.56}$ & ${40.87}$ \\
  &  &    ${-1.079}$  & ${19.46}$ & ${-21.36}$ & 
${154.46}$ & ${57.58}$ \\
A1689 & $V$ &   $-$1.500 &  18.83 & $-$21.99 & 57.25 & 33.33 & 3.70 \\
& & ${-1.382}$  & ${19.00}$ & ${-21.82}$ &  ${85.73}$ & ${30.48}$ \\
  &  &    ${-1.647}$  & ${18.56}$ & ${-22.26}$ & 
${30.93}$ & ${34.92}$ \\
\end{tabular}
\label{tab:fieldins}
\end{table*}

\begin{table*}
\centering
\caption{Schechter function parameter fits for incompleteness
and obscuration corrected field subtracted points. The column headings
are as in Table~\ref{tab:fieldins}.}
\vspace{0.5cm}
\begin{tabular}{lccccrrc}
{Cluster} & {Filter} & $\alpha$ & $m^{\star}$ & $M^{\star}$ &
$N^{\star}$ & $\chi^{2}$ & $\chi^{2}$/$\nu$  \\
A665  & $I$ &    $-$1.188  & 18.04 & $-$22.46& 123.00 & 26.03 & 2.89 \\
& & 
${-1.185}$  & ${17.98}$ & ${-22.52}$ & 
${139.02}$ & ${25.87}$ \\
& & 
${-1.222}$  & ${18.04}$ & ${-22.46}$ & 
${95.77}$ & ${26.69}$ \\
A1689 & $I$ &   $-$1.430  & 17.30 & $-$23.20& 57.04 & 22.12 & 2.12\\
& & 
${-1.406}$  & ${17.28}$ & ${-23.22}$ & 
${66.52}$ & ${23.01}$ \\
& & 
${-1.466}$  & ${17.37}$ & ${-23.13}$ & 
${47.38}$ & ${21.42}$ \\
A665  & $V$ &    $-$1.167  & 19.37 & $-$21.45& 178.36  & 59.17 & 6.57\\
& & 
${-1.042}$  & ${19.43}$ & ${-21.39}$ & 
${233.79}$ & ${52.01}$ \\
& & 
${-1.370}$  & ${19.00}$ & ${-21.82}$ & 
${96.92}$ & ${64.57}$ \\
A1689 & $V$ &    $-$1.607  & 18.68 & $-$22.14& 51.40 & 35.96 & 4.00\\
& &
${-1.514}$  & ${18.84}$ & ${-21.98}$ & 
${74.56}$ & ${34.07}$ \\
& &
${-1.715}$  & ${18.48}$ & ${-22.34}$ & 
${31.70}$ & ${36.80}$ \\
\end{tabular}
\label{tab:fieldinars}
\end{table*}

\begin{table*}
\centering
\caption{Schechter function parameter fits for incompleteness and
obscuration corrected radial inner/outer test points. The column headings
are as in Table~\ref{tab:fieldins}.}
\vspace{0.5cm}
\begin{tabular}{lccccrcc}
{Cluster} & {Filter} & $\alpha$ & $m^{\star}$ & $M^{\star}$ & 
$N^{\star}$ & $\chi^{2}$ & $\chi^{2}$/$\nu$  \\
A665  & $I$ &    $-$1.235 &    17.49 & $-$23.01& 40.96& 1.27  & 0.14\\
A1689 & $I$ &   $-$1.034 &  17.43   & $-$23.07&139.84 &10.24   &1.02 \\
A665  & $V$ &    $-$1.264 &  18.82  & $-$22.00& 58.07 & 4.42 & 0.49 \\
A1689 & $V$ &   $-$1.681  &  18.09  & $-$22.73&19.10 &5.67& 0.63\\
\end{tabular}
\label{tab:ioinars}
\end{table*}

\begin{table*}
\centering
\caption{Gaussian+Schechter function parameter fits.  The top table
gives the fits for the incompleteness and obscuration corrected
catalogues, while the lower is for the incompleteness corrected
case only.  The columns show the cluster, filter, apparent magnitude at the
peak of Gaussian function, absolute magnitude at the peak of Gaussian
function, normalisation of Schechter function, faint end slope of
Schechter function, apparent magnitude at the knee of Schechter
function, absolute magnitude at the knee of Schechter function,
$\chi^{2}$, and reduced $\chi^{2}$ values. Note that $\sigma$, the
standard deviation of the Gaussian function  is held constant at 1.0
and $K$, the normalisation of the Gaussian function is required to
equal $0.1\times N^{\star}$.}
\vspace{0.5cm}
\begin{tabular}{lcccccccrc}
{Cluster} & {Filter}  &$m^{\rm cent}$& $M^{\rm cent}$ &$N^{\star}$  &$\alpha$ 
& $m^{\star}$ & $M^{\star}$ &$\chi^{2}$ & $\chi^{2}$/$\nu$ \\
A665 & $I$ & 18.36 & $-$22.14   &312.21 & $-$0.991  &  20.01 &$-$20.49& 
12.01& 2.00 \\
A1689& $I$ & 18.44 & $-$22.06   &319.99 & $-$1.180  &  20.03 &$-$20.47& 
11.89& 1.49 \\
Combined & $I$ &18.66 & $-$21.84   &343.98 & $-$1.160  &  20.23 &$-$20.27 &
12.32& 1.76 \\
A665 & $V$ &20.55 & $-$20.27   &823.73 & $-$2.295  &  22.69 &$-$18.13 &
9.51& 1.36 \\
A1689& $V$ &20.63 & $-$20.19  &689.95 & $-$1.870  &  22.32 &$-$18.50 &
9.99& 1.43 \\
Combined & $V$ &20.62 & $-$20.20   &776.95 & $-$2.092  &  22.55 &$-$18.27 &
14.68& 2.10 \\
\end{tabular}
\begin{tabular}{lcccccccrc}
{Cluster} & {Filter}  &$m^{\rm cent}$& $M^{\rm cent}$ &$N^{\star}$  &$\alpha$ 
& $m^{\star}$ & $M^{\star}$ &$\chi^{2}$ & $\chi^{2}$/$\nu$ \\
A665 & $I$ & 18.22 & $-$22.28   &268.05 & $-$0.841  &  19.97 &$-$20.53& 
11.68& 1.95 \\
A1689& $I$ & 18.42 & $-$22.08   &290.07 & $-$1.025  &  20.18 &$-$20.32& 
12.01& 1.50 \\
Combined & $I$ &18.64 & $-$21.86   &309.18 & $-$1.017  &  20.36 &$-$20.23 &
12.31& 1.76 \\
A665 & $V$ &20.48 & $-$20.34   &673.15 & $-$2.464  &  22.81 &$-$18.01 &
9.17& 1.31 \\
A1689& $V$ &20.60 & $-$20.22  &572.42 & $-$1.956  &  22.45 &$-$18.37 &
8.92& 1.27 \\
Combined & $V$ &20.56 & $-$20.26  &635.45 & $-$2.230  &  22.66 &$-$18.16 &
12.69& 1.81\\
\end{tabular}

\label{tab:gs}
\end{table*}

\section*{ACKNOWLEDGMENTS}

We are grateful to Chris Lidman for his generosity in providing and
analysing the field survey, without which our analysis would have been
impossible.  We also thank Chris for answering all of our naive
questions about his dataset.  We thank Simon Driver, Neil Trentham and
Ann Zabludoff for useful discussions.  Many thanks are also due to
Carlos Frenk for his help, support and encouragement on this project.
GW gratefully acknowledges a PPARC studentship during the time these
observations were made. Support via a Carnegie Fellowship is gratefully
acknowledged by IRS. IRS and RSE acknowledge support from PPARC.  WJC
acknowledges support from the Australian Department of Industry,
Science and Technology, the Australian Research Council and Sun
Microsystems. The observations described in this paper were obtained at
the Isaac Newton and William Herschel Telescopes, Observatorio del
Roque de los Muchachos, La Palma.

\bibliography{lensing,lumfn}

\bibliographystyle{mnras}

\bsp

\end{document}